# Term-based composition of security protocols

B. Genge[1], P. Haller[1], R. Ovidiu[1], I. Ignat[2]

[1]"Petru Maior" University of Targu Mures, Romania, bgenge@upm.ro, phaller@upm.ro, oratoi@engineering.upm.ro
[2]Technical University of Cluj Napoca, Romania, Iosif.Ignat@cs.utcluj.ro

*Abstract*-In the context of security protocol parallel composition, where messages belonging to different protocols can intersect each other, we introduce a new paradigm: term-based composition (i.e. the composition of message components also known as terms). First, we create a protocol specification model by extending the original strand spaces. Then, we provide a term composition algorithm based on which new terms can be constructed. To ensure that security properties are maintained, we introduce the concept of term connections to express the existing connections between terms and encryption contexts. We illustrate the proposed composition process by using two existing protocols.

## I. INTRODUCTION

Security protocols are communication protocols in which participants use encryption to send each other encoded information. With the rapid growth of the Internet and a desperate need to secure communication, in the last few decades the attention of many researchers has been led towards the analysis of security protocols [1], [2], [3], [4], [5], [6].

Recently, there have been several proposals developed to help the process of security protocol design using formal methods and tools [7], [8], [9], [10], [11], [12], [13]. Most of the proposed techniques use a modular approach in the design process, where the user is given a set of small protocols from which more complex protocols can be constructed, process also known as *composition* [9], [10], [11].

In the existing composition techniques, authors mainly deal with the sequential and parallel composition of security properties viewed as a set of information transmitted over messages. However, the composition of message components has not been addressed in a proper manner, meaning that users have to solve the problem of creating new messages on their own.

Solving this problem, apparently insignificant, can lead to protocols which execute in half the time the original, composed protocols do. In addition, the composition process can lead to multiple results, which must be carefully analyzed on a message level to increase protocol performance.

In this paper, we introduce a novel composition paradigm: *term-based composition*. The composition problem is addressed at the message level based on syntactical constructions and analysis. This new paradigm is addressed in the context of *parallel composition*, where protocol messages intersect each other. The resulting protocol contains not only a set of unified messages but also a unified set of security properties (e.g. secrecy, authentication, integrity).

The paper is structured as follows. Section II introduces the concept of k-strands used to model security protocols. Security requirements are addressed in section III. In section IV we present the problem of generating protocols using parallel composition and term-based composition and we propose a term composition algorithm. We exemplify the composition process by composing two protocols.

## II. KNOWLEDGE STRANDS

In this section we briefly present the concept of *knowledge strands* (*k-strands*). For a more detailed presentation, the reader is directed to consult the authors' previous work [6], [17].

A *strand* is a sequence of transmission and reception events used to model protocol participants. A collection of strands is called a *strand space*. The strand space model was introduced by Fabrega, Herzog and Guttman in [15] and extended by the authors with participant knowledge, specialized basic sets and explicit *term* construction in [5], [6]. The resulting model is called a *k-strand space*. The rest of this section formally defines the k-strand and k-strand space concepts.

By analyzing the protocol specifications from the SPORE library [20] we can conclude that protocol participants communicate by exchanging *terms* constructed from elements belonging to the following sets: R, denoting the set of participant names; N, denoting the set of nonces (i.e. "number once used") and K, denoting the set of cryptographic keys. If required, other sets can be easily added without affecting the other components.

To denote the encryption type used to create cryptographic terms, we define the following *function names*:

$$FuncName ::= \text{sk} \quad (secret\ key) \quad (1)$$
$$| \text{pk} \quad (public\ key)$$
$$| \text{pvk} \quad (private\ key)$$
$$| \text{h} \quad (hash).$$

The above-defined basic sets and function names are used in the definition of *terms*, where we also introduce constructors for pairing and encryption:

$$\mathcal{T} ::= . | \mathsf{R} | \mathsf{N} | \mathsf{K} | (\mathcal{T},\mathcal{T}) | \{\mathcal{T}\}_{FuncName(\mathcal{T})}, \quad (2)$$

where the '.' symbol is used to denote an empty term. We use the symbol $\mathcal{T}^*$ to denote the set of all subsets of terms.

The composition process of two terms $t_1$ and $t_2$ into another term $t$ implies that $t$ has sub-terms. The sub-term relation $\prec$ is inductively defined as follows.

**Definition 1.** *The sub-term relation $\prec$ is the smallest relation on terms such that:*
1. $t \prec t$;
2. $t \prec \{t_1\}_{f(t_2)}$ *if* $t \prec t_1$ *or* $t \prec t_2$;
3. $t \prec (t_1, t_2)$ *if* $t \prec t_1$ *or* $t \prec t_2$.

Before defining the concept of knowledge strands we need to define another element: *classifiers*. As suggested by their names, classifiers are used to classify or categorize knowledge strands. The categories are created based on the type of operation modeled by a given knowledge strand. Formally, classifiers are defined as:

$$C ::= C_\mathcal{R} \quad (Participant\ classifier) \quad (3)$$
$$|\ C_\mathcal{M} \quad (Memory\ classifier)$$

To denote the transmission and reception of terms, we use *signed terms*. The occurrence of a term with a positive sign denotes transmission, while the occurrence of a term with a negative sign denotes reception. The set of transmission and reception sequences is denoted by $(\pm \mathcal{T})^*$.

**Definition 2.** *A k-strand (i.e. knowledge strand) is a tuple $\langle \mathcal{K}, c, r, s \rangle$, where $\mathcal{K} \in \mathcal{T}^*$ denotes the knowledge corresponding to the modeled participant, $c \in C$ denotes the classifier, $r \in \mathsf{R}$ denotes the participant name and $s \in (\pm \mathcal{T})^*$ denotes the sequence of transmissions and receptions. A set of k-strands is called a k-strand space. The set of all k-strand spaces is denoted by $\Sigma_k$. Let $\varsigma_k$ be a k-strand space and $s_k \in \varsigma_k$ a k-strand, then:*

1. *We define the following mapping functions: $kknow(s_k)$ to map the knowledge component; $kclass(s_k)$ to map the classifier component; $kpart(s_k)$ to map the name component; $kstrand(s_k)$ to map the term sequence component;*
2. *A node is any transmission or reception of a term, written as $n = \langle kstrand(s_k), i \rangle$, where $i$ is an integer satisfying the condition $1 \le i \le length(s)$. We define the $term(n)$ function to map the term corresponding to a given node;*
3. *Let $n_1 = \langle kstrand(s_k), i \rangle$ and $n_2 = \langle kstrand(s_k), i+1 \rangle$ be two consecutive nodes from the same k-strand. Then, there exists an edge $n_1 \Rightarrow n_2$ in the same k-strand;*
4. *Let $n_1, n_2$ be two nodes. If $n_1$ is a positive node and $n_1$ is a negative node belonging to different k-strands, then there exists an edge $n_1 \to n_2$. We define the $sign(n)$ function to map the sign of a given node.*

Fig. 1 shows an example specification of Lowe's BAN Concrete Secure RPC [14] protocol in the described k-strand space model.

III. SECURITY REQUIREMENTS

The composition of security protocols can not be made by simply adding messages to one protocol. By inspecting the rather large number of reported attacks in the literature [14], [18], [20] we can agree that any modification brought upon a protocol can influence its existing security properties. Based on these concerns, the authors have developed in a previous paper [17] a framework for verifying the composability of security protocols.

The method developed by the authors requires the execution of two steps. First, we must verify if secret terms from one protocol can be found in *insecure* terms in the other protocol. By the concept *insecure* we mean terms encrypted with insecure keys (e.g. session keys) or terms that are sent out clearly. Second, we must verify if terms encrypted with the same key are structurally independent. In other words, we must verify if participants, based on term structures and knowledge can distinguish between the given terms.

The first requirement is fulfilled by conducting a syntactical verification of the given protocol terms. The protocol model used is the one presented in the previous section. Alongside the specification, the user has to provide the terms considered to be secret for each protocol.

For the second requirement to be fulfilled we must construct the canonical specification model proposed by the authors in the same paper. This model eliminates instantiation-based information (e.g. Na, A, B, Kab), leaving only essential information needed in the structural independence verification process (e.g. n, r, r, k).

IV. COMPOSITION

*A. Generating protocols*

By using parallel composition, we can produce several distinct protocols. For example, given two protocols, P1 and P2, each of them with two messages, the protocols that can be constructed are listed in Table 1, where P1.i and P2.j, $i, j \in \{1, 2\}$, denote message indexes corresponding to the two protocols and Px.i,Py.j, $x, y \in \{1, 2\}$ denotes concatenation.

TABLE I
PROTOCOL AND MESSAGE SEQUENCES GENERATED
USING PARALLEL MESSAGE COMPOSITION

| Without term composition | | With term composition | |
|---|---|---|---|
| P1.1<br>P1.2<br>P2.1<br>P2.2 | P1.1<br>P2.1<br>P1.2<br>P2.2 | P2.1, P1.1<br>P1.2<br>P2.2 | P1.1<br>P2.1, P1.2<br>P2.2 |
| P2.1<br>P1.1<br>P2.2<br>P1.2 | P1.1<br>P2.1<br>P2.2<br>P1.2 | P2.1, P1.1<br>P2.2, P1.2 | P1.1<br>P2.1<br>P2.2, P1.2 |
| P2.1<br>P2.2<br>P1.1<br>P1.2 | P2.1<br>P1.1<br>P1.2<br>P2.2 | P2.1<br>P1.1<br>P2.2, P1.2 | P2.1, P1.1<br>P2.2<br>P1.2 |
| - | - | P2.1<br>P2.2, P1.1<br>P1.2 | - |

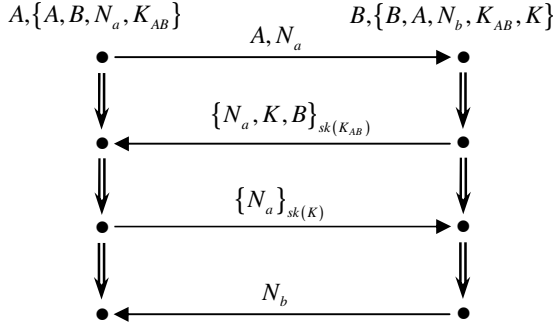

Figure 1. Lowe's BAN Concrete Andrew
Secure RPC representation in the k-strand space model

More formally, given two protocols modeled in the k-strand space, $\varsigma_k, \varsigma'_k \in \Sigma_k$, we generate new protocols using operations such as *message intercalation* and *term concatenation*. *Message intercalation* denotes the process by which several messages belonging to different protocols are combined together maintainig at the same time their original order of appearance. On the other hand, *term concatenation* simply concatenates two terms without performing any optimisations on the resulting term.

The generated protocols are denoted by the set *GenProtPairs*. Each element of this set contains a sequence of term pairs $\langle x_i, y_j \rangle$, where the first component denotes terms transmitted in the first protocol and the second component denotes terms transmitted in the second protocol. More formally, $x_i \in sentTerms(\varsigma_k)$, $y_j \in sentTerms(\varsigma'_k)$, $i = \overline{1, |sentTerms(\varsigma_k)|}$, $j = \overline{1, |sentTerms(\varsigma'_k)|}$, where $sentTerms : \Sigma_k \to \mathcal{T}^*$ is a function mapping the set of sent terms in a given protocol specification, defined as:

$$sentTerms(\varsigma_k) = \left\{ \bigcup_{\substack{s_k \in \varsigma_k \\ n_i = \langle kstrand(s_k), i \rangle, sign(n_i) = +}}^{i=1, length(kstrand(s_k))} term(n_i) \right\} \cup \{.\} \quad (4)$$

This function also mapps empty components, denoted by '.' to model sittuations where the second operation (i.e. term concatenation) is not applied.

As a final step for the protocol generation process, we must check that concatenated messages have the same source and destination participants. If we find at least one message that does not satisfy this requirement, the entire protocol is removed from the list.

*B. Security property definition*

The term composition process constructs all possible combinations of terms using two given terms by modifying existing terms. In the context of security protocols, these combinations must not destroy existing security properties. In order to provide a correct composition we must define the concept of a security property.

Because security protocols consist of participants exchanging terms, security properties are created by the transmitted and received terms. More specifically, it is the cryptographic context of each term in conjunction with the exchange of terms from which security properties are constructed. To formally define security properties we introduce two new concepts: *partial* and *complete term connection*. *Connections* between terms denote the existence of a set of common terms. *Partial connections* denote the connections between a free (i.e. unencrypted) term and an encrypted one while *complete connections* denote the connections between two encrypted terms.

To express the existence of a partial and a complete connection we introduce two operators, $\_ \mapsto_P \_ : \langle (\pm \mathcal{T}) \times \mathcal{T} \times \Sigma_k \rangle \times \langle (\pm \mathcal{T}) \times \mathcal{T} \times \Sigma_k \rangle$ and $\_ \mapsto_C \_ : \langle (\pm \mathcal{T}) \times \mathcal{T} \times \Sigma_k \rangle \times \langle (\pm \mathcal{T}) \times \mathcal{T} \times \Sigma_k \rangle$ respectively. These operators denote the connection between one node, term and k-strand to another node, term and k-strand. The first component of these operators is called a *pre-condition* and the second is called a *post-condition*. We define the following functions, *cnode, cterm, cstrand* to map the node, term and k-strand corresponding to a pre-condition or post-condition.

We say that there is a partial connection between two terms $t_1$ and $t_2$, if $t_1$ is a sub-term of $t_2$, $t_1$ is not encrypted and $t_2$ has a cryptographic construction or viceversa. Formally,

$$\langle n_1, t_1, s_{k1} \rangle \mapsto_P \langle n_2, t_2, s_{k2} \rangle \text{ if } t_1 \prec t_2, \text{ where} \quad (5)$$

$$\left( t_1 \neq \{t'_1\}_{f(t)} \wedge t_2 = \{t'_2\}_{f(t)} \right) \vee$$
$$\left( t_1 = \{t'_1\}_{f(t)} \wedge t_2 \neq \{t'_2\}_{f(t)} \right),$$
$$t_1 \prec term(n_1), t_1 \prec term(n_1),$$
$$n_1 \in \{n_i \mid 1 \leq i \leq length(kstrand(s_{k1}))\},$$
$$n_2 \in \{n_i \mid 1 \leq i \leq length(kstrand(s_{k2}))\}.$$

A complete connection between two terms, $t_1$ and $t_2$, exists only if $t_1$ is an encrypted sub-term of $t_2$ and $t_2$ has a cryptographic construction or the non-cryptographic component of $t_1$ is a sub-term of $t_2$. Formally,

$$\langle n_1, t_1, s_{k1} \rangle \mapsto_C \langle n_2, t_2, s_{k2} \rangle \text{ if } t_1 \prec t_2 \text{ or } t'_1 \prec t_2, \text{ where} \quad (6)$$

$$t_1 = \{t'_1\}_{f(t)}, \ t_2 = \{t'_2\}_{f(t)},$$
$$t_1 \prec term(n_1), t_1 \prec term(n_1),$$
$$n_1 \in \{n_i \mid 1 \leq i \leq length(kstrand(s_{k1}))\},$$
$$n_2 \in \{n_i \mid 1 \leq i \leq length(kstrand(s_{k2}))\}.$$

**Definition 3.** *A security property $\xi$ is a collection of partial and complete connections.*

By the definition given above, a security property is a set of connections between terms. This definition is similar to the definition of authenticaton tests given by Guttman in [10]. The difference is that we define connections not only between terms transmitted by different nodes, but also between sub-terms. This allows us to define complex security properties such as authentication, but also other, more subtle ones such as secrecy.

By using term connections we can model dependencies between terms. This key aspect is vital in the process of term composition because by modifying one term we must also modify other, dependent terms to maintain existing security properties.

*C. Modeling dynamic knowledge*

As opposed to the *static* (i.e. initial) knowledge, there is another type of knowledge that can be constructed by protocol participants: *dynamic knowledge*. This type of knowledge grows with every term that is received. Dynamic knowledge is modeled as a k-strand that "communicates" with the participant's k-strand using term transmissions and receptions.

Participants are modeled as a pair of k-strands consisting of one main, participant k-strand and a memory k-strand, modeling dynamic knowledge. In the composition process, terms can be modified. For example, they can be included in cryptographic context that can not be created by a participant because at the given node cryptographic keys have not yet been received. By modeling dynamic knowledge, we are able to decide if the terms that must be transmitted by a node can be constructed.

In order to provide a persistent model of the dynamic knowledge, we consider that terms from this knowledge are stored in a memory region that can only be accessed by the corresponding participant. This memory region, as mentioned earlier, is modeled as a k-strand. However, because communication between each participant and its attached memory must be private, we consider an encrypted communication model using a new function type *mk* and a key. The function is the same, while the key is unique for each user.

Next, we propose an algorithm for creating memory k-strands, identified by the class $C_\mathcal{M}$. Given an initial k-strand $s_k$, that models the operations corresponding to a participant, by running the algorithm, we generate two new k-strands, a participant k-strand $s'_k$ and a memory k-strand $s''_k$. The newly generated participant k-strand additionally contains nodes modeling communication with the attached memory k-strand.

Receiving a term from the memory k-strand corresponds to the dynamic knowledge. The terms received by memory k-strands are decoded, transformed into new knowledge and added to the existing knowledge.

The proposed algorithm makes use of the $genKnow: \mathcal{T} \times \mathcal{T} \to \mathcal{T}$ function to generate new knowledge based on existing knowledge (stored as a term) and a new received term.

Algorithm 1. *Memory k-strand generation:*
1. Generate memory communication encryption key $K_m$
2. Initialize the new k-strands:
$$s'_k = \langle \{kknow(s_k) \cup K_m\}, C_\mathcal{R}, r, \langle . \rangle \rangle$$
$$s''_k = \langle \{kknow(s_k) \cup K_m\}, C_\mathcal{M}, r, \langle . \rangle \rangle$$

2. For every positive node $n = \langle kstrand(s_k), i \rangle$ add a positive node to $s'_k$:
$$s'_k = \langle kknow(s'_k), C_\mathcal{R}, r, \langle kstrand(s'_k), n \rangle \rangle$$

3. For every negative node $n = \langle kstrand(s_k), i \rangle$ add a negative node to $s'_k$ and generate new knowledge:

$$s'_k = \langle kknow(s'_k), C_\mathcal{R}, r, \langle kstrand(s'_k), n, +\{term(n)\}_{mk(K_m)} \rangle \rangle$$

$$s''_k = \langle kknow(s''_k), C_\mathcal{M}, r, \langle kstrand(s''_k), -\{term(n)\}_{mk(K_m)} \rangle \rangle$$

Let $n'$ be the last positive node from $s''_k$
Let $\mathcal{K}' = kknow(s'_k)$ and $\mathcal{K}'' = kknow(s''_k)$
Let $t_{know} = genKnow(term(n), term(n'))$

$$s'_k = \langle \mathcal{K}', C_\mathcal{R}, r, \langle kstrand(s'_k), -\{t_{know}\}_{mk(K_m)} \rangle \rangle$$

$$s''_k = \langle \mathcal{K}'', C_\mathcal{M}, r, \langle kstrand(s''_k), +\{t_{know}\}_{mk(K_m)} \rangle \rangle$$

*D. Term composition algorithm*

In the protocol generation process described at section *A*, terms that are concatenated must be composed in order to generate more performant protocols. The composition process can alter terms, maintaining at the same time existing security properties.

First, we construct the connection sequences between protocol terms for the involved protocols. Then, we initialize a new k-strand space by creating k-strands corresponding to participants. The initialization process also creates unified static knowledge sets for every participant.

Next, for every pair of concatenated terms resulted in the protocol generation phase we run the composition algorithm. By modifying one term we must ensure that the terms from the connection sequence are also modified. We ensure that partial connections are maintained by not modifying the cryptographic context of terms. Maintaining complete connections, however, requires a subsequent modification of dependent terms.

After performing each term composition, the memory k-strand algorithm from section *C* is run to construct the memory k-strands. Then, for every term transmitted by a participant k-strand we use the $Constructable: \mathcal{T} \times \mathcal{T}^* \times \mathcal{T}$ predicate to verify if the transmitted term can be constructed from the existing static and dynamic knowledge.

For two concatenated terms $(t_1, t_2)$, $t_1 \in sentTerms(\varsigma'_k)$, $t_2 \in sentTerms(\varsigma''_k)$, the composition algorithm is the following.

Algorithm 2. *Composition:*
1. Construct connection sequences as security properties:
Let $\xi' = \{\langle n_1, t_1, s_{k1} \rangle \mapsto_{C,P} \langle n_2, t_2, s_{k2} \rangle, | s_{k1}, s_{k1} \in \varsigma'_k\}$
Let $\xi'' = \{\langle n_1, t_1, s_{k1} \rangle \mapsto_{C,P} \langle n_2, t_2, s_{k2} \rangle, | s_{k1}, s_{k1} \in \varsigma''_k\}$

2. Initialize new k-strand space:
Let $\varsigma_k$ be the resulting k-strand space
For each $s'_k \in \varsigma'_k \cup \varsigma''_k$ do
   If $(\forall s_k \in \varsigma_k, krole(s_k) \diamond krole(s'_k))$ then
     Let $s_k = \langle kknow(s_k), krole(s_k), C_\mathcal{R}, \langle . \rangle \rangle$
     $\varsigma_k = \varsigma_k \cup \{s_k\}$
   Else
     Let $s_k \in \varsigma_k : krole(s_k) = krole(s'_k)$ and
$$s_k = \langle \mathcal{K}, r, c, s \rangle$$

$s_k = \langle \{\mathcal{K} \cup kknow(s'_k)\}, r, c, s \rangle$

   EndIf
EndFor
3. Compose two terms:
   Let $t_1 = \{t'_1\}_{f_1(k_1)}$, $t_2 = \{t'_2\}_{f_2(k_2)}$,
   $t_1 \prec term(n_1), t_2 \prec term(n_2)$
   If $f_1 = f_2 \wedge k_1 = k_2$ then
     If $\nexists (c_1 \mapsto_C c_2 \in \xi') : (cterm(c_2) = t'_1 \wedge cnode(c_2) = n_1)$
        $\vee (cterm(c_1) = t'_1 \wedge cnode(c_1) = n_1)$ then
        $t_1 = \{t'_1, t''_2\}_{f_1(k_1)}$
     Else
     If $\nexists (c_1 \mapsto_C c_2 \in \xi')$:
        $(cterm(c_1) = t'_1 \wedge cnode(c_1) = n_1)$ then
        $t_1 = \{t'_1, t''_2\}_{f_1(k_1)}$
        @update term connection sequence
     EndIf
   Else
     $t_1 = \{t'_1, t_2\}_{f_1(k_1)}$
     @update term connection sequence
   EndIf
4. Generate memory k-strands
   @run Algorithm 1 to construct $\varsigma_k$ initialized at step 2
5. Verify term generation
   Let $s_k, s'_k \in \varsigma_k : kclass(s_k) = C_\mathcal{R} \wedge kclass(s'_k) = C_\mathcal{M} \wedge$
   $kpart(s_k) = kpart(s'_k)$
   Let $n$, $n'$ be the last positive node from $s_k$ and $s'_k$ respectively
   If $Constructable(term(n), kknow(s_k), term(n'))$ then
     @Accept $\varsigma_k$
   Else
     @Reject $\varsigma_k$
   EndIf

## V. COMPOSITION EXAMPLE

To illustrate the composition process we use two protocols: "Woo and Lam Pi3" [16] and Lowe's modified version of the Yahalom [18, 19] protocol. The k-strand representation of the two protocols can be seen in Fig. 2 and Fig. 3. We use $\varsigma_k$ to model the k-strand space corresponding to the "Woo and Lam Pi3" and $\varsigma'_k$ to model the k-strand space corresponding to Lowe's Yahalom protocol.

The first step towards the composition of these protocols consists in verifying the „key-secrecy independence" security requirement formulated by the authors in [17]. To achieve this, we specify the secret terms for the two involved protocols. For the first protocol, these are no secret terms, while the secret terms for the second protocol are $\{N_b, K_{AB}\}$ (we consider that participant names are public).

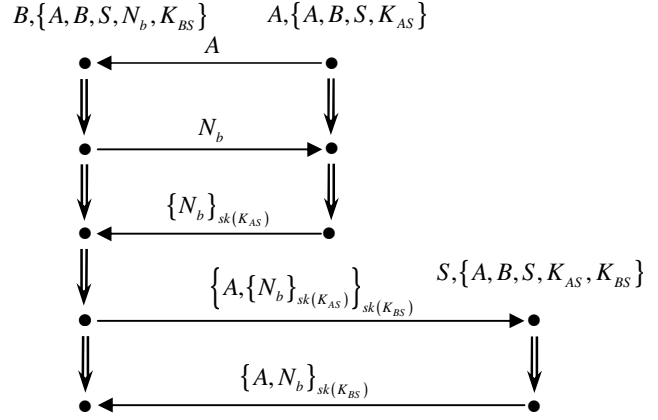

Figure 2. Woo and Lam Pi3 representation in the k-strand space model

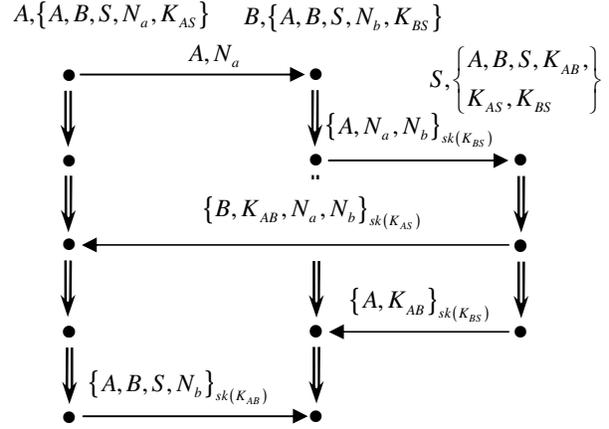

Figure 3. Lowe's modified version of Yahalom's representation in the k-strand space model

Because $\exists t \in sentTerms(\varsigma_k) : N_b \prec t$ and $t$ is not encrypted, the first requirement is not satisfied. To allow the composition of the two protocols, $N_b$ in the first protocol must be different from $N_b$ in the second protocol. We emphasize this aspect by replacing $N_b$ with $N'_b$ in $\varsigma'_k$.

Because of space considerations, we only construct complete connections which play a crucial role in the composition process. In protocol $\varsigma_k$ we have only one complete connection:

$\langle +\{N_b\}_{sk(K_{AS})}, \{N_b\}_{sk(K_{AS})}, s_{kA} \rangle \mapsto_C$
$\qquad \langle +\{N_b\}_{sk(K_{AS})}, \{A, \{N_b\}_{sk(K_{AS})}\}_{sk(K_{AS})}, s_{kB} \rangle$

Because of term structure varieties, in protocol $\varsigma'_k$ there are no complete connections. By using the steps described in section IV.B we generate all possible sequences of protocols, resulting a total number of 1683 protocols. After filtering protocols for which concatenated terms have different source-destination participants, there remain a total number of 408 protocols. For each protocol we can apply the term composition algorithm, resulting a new set of protocols. One of the resulting protocols is shown in Fig. 4. In order to select the most performant protocols, we can apply the „minimum number of messages" principle,

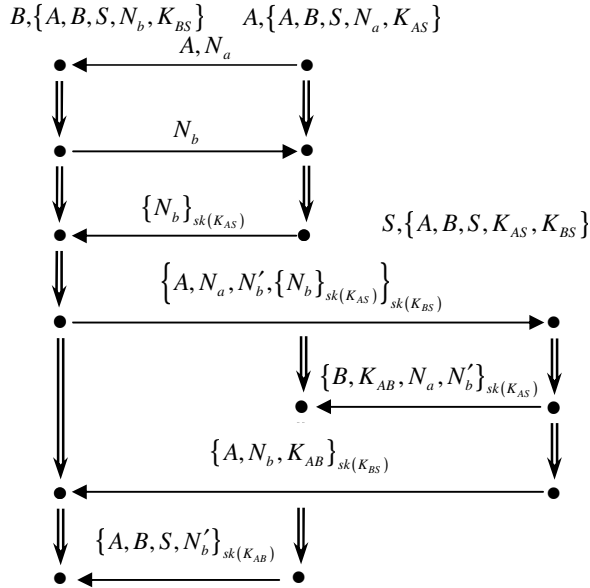

Figure 4. Composed protocol

or we can construct a performance evaluation method, which we consider to be part of future work.

As we can see from Fig. 4, the complete connection is also maintained in the composed protocol. In addition, the second security requirement formulated by the authors in [17], i.e. "message independence", is also satisfied because messages have different cryptographical structures.

## VI. CONCLUSIONS AND FUTURE WORK

In this paper we proposed a method for composing security protocol terms. To define security properties embedded in protocols we introduced the concept of partial and complete connections. Our approach modifies terms only in the sense of extending them with new components, thus preserving partial connections. Complete connections are maintained by modifying all subsequent terms dependent of the modified term.

As future work, we intend to extend the proposed term composition algorithm with performance-related information. This would give users the possibility to choose the best suited protocol for a given environment. However, this is rather difficult to achieve based only on informal specifications. This is why we intend to construct a performance evaluation model that allows us to compare protocol performance rather than giving an exact behavior in a specific environment.